\documentclass[aps, prx,twocolumn,longbibliography,superscriptaddress,amsmath,amssymb,floatfix]{revtex4}
\usepackage[latin9]{inputenc}
\usepackage{amssymb}
\usepackage{graphicx}
\usepackage{amsmath}
\usepackage{color}
\usepackage{mathrsfs}
\usepackage{float}
\usepackage{indentfirst}
\usepackage{mathrsfs}
\usepackage{float}
\usepackage{indentfirst}
\usepackage{textcomp}
\usepackage{comment}
\usepackage{mathtools}
\usepackage{natbib,hyperref}
\usepackage{soul}

\begin{document}

\title{Comparative study of charge order in undoped infinite-layer
nickelate superconductors}
\author{Yang Shen}
\affiliation{Key Laboratory of Artificial Structures and Quantum Control, School of
Physics and Astronomy, Shanghai Jiao Tong University, Shanghai 200240, China}
\author{Mingpu Qin}
\thanks{qinmingpu@sjtu.edu.cn}
\affiliation{Key Laboratory of Artificial Structures and Quantum Control, School of
Physics and Astronomy, Shanghai Jiao Tong University, Shanghai 200240, China}
\author{Guang-Ming Zhang}
\thanks{gmzhang@tsinghua.edu.cn}
\affiliation{State Key Laboratory of Low-Dimensional Quantum Physics and Department of
Physics, Tsinghua University, Beijing 100084, China}
\affiliation{Frontier Science Center for Quantum Information, Beijing 100084, China}

\begin{abstract}
To understand the microscopic mechanism of the charge order observed in the
parent compound of the infinite-layer nickelate superconductors, we consider
a minimal three-legged model consisting of a two-legged Hubbard ladder for
the Ni $3d_{x^2-y^2}$ electrons and a free conduction electron chain from
the rare-earths. With highly accurate density matrix renormalization group
calculations, when the chemical potential difference is adjusted to make the
Hubbard ladder with $1/3$ hole doping, we find a long-range charge order
with period $3$ in the ground state of the model, while the spin excitation
has a small energy gap. Moreover, the electron pair-pair correlation has a
quasi-long-range behavior, indicating an instability of superconductivity
even at half-filling. As a comparison, the same method is applied to a pure
two-legged Hubbard model with $1/3$ hole doping in which the period-3 charge
order is a quasi-long range one. The difference between them demonstrates
that the free electron chain of the three-legged ladder plays the role of a
charge reservoir and enhances the charge order in the undoped infinite-layer nickelates.
\end{abstract}

\maketitle
\section{Introduction}
Since 2019, the family of infinite-layer nickelate superconductors has been
synthesized\cite%
{li2019superconductivity,nomura2022superconductivity,PhysRevLett.125.027001,PhysRevLett.125.147003,PhysRevMaterials.4.121801,https://doi.org/10.1002/adma.202104083,doi:10.1126/sciadv.abl9927}
which provides another useful platform for the exploration of the
microscopic mechanism of unconventional superconductivity. Similar to the
cuprates, the infinite-layer nickelates with nominal $3d^9$ $Ni^+$ contain $%
3d_{x^2-y^2}$ orbital degree of freedom on a quasi-two-dimensional Ni-O
square lattice near the half-filling\cite%
{anisimov1999electronic,lee2004infinite}. However, due to the much larger $p$%
-$d$ energy splitting, their low-energy electronic structures are more
likely to fall into the Mott-Hubbard than the charge-transfer regime in the
Zaanen-Sawatzky-Allen classification scheme\cite%
{PhysRevLett.55.418,doi:10.1073/pnas.2007683118}. Different from an
antiferromagnetic Mott insulator, the resistivity of the parent compounds
exhibits metallic behavior at high temperatures and an upturn at low
temperatures\cite{li2019superconductivity,zhang-yang-zhang2020}. No static
long-range magnetic order are observed down to the lowest measured
temperature\cite{Hayward-SSS2003} though spin fluctuations are reported in
experiments\cite{Lu-Science2021}.

Very recently, a vertical charge order (CO) with period $3$ was reported in
the undoped parent compound of the infinite-layer nickelates\cite%
{rossi2021broken,krieger2021charge,tam2021charge}. The vertical CO is along
the direction of Ni-O bond, different from the diagonal stripe order found
in other nickelates\cite{PhysRevLett.73.1003} where superconductivity is
absent. The CO becomes weaker with the increase of doping and finally
disappears at $20\%$ doping where superconductivity sets in. Moreover, the
X-ray absorption spectroscopy experiments have shown that the doped holes
are mainly localized in the Ni $3d_{x^2-y^2}$ orbital, playing the dominant
role in the low-energy physics\cite{Rossi-prb2021}. So it is highly
desirable to understand the nature of the undoped parent compound of nickelates.

\begin{figure}[tbp]
\includegraphics[width=0.45\textwidth]{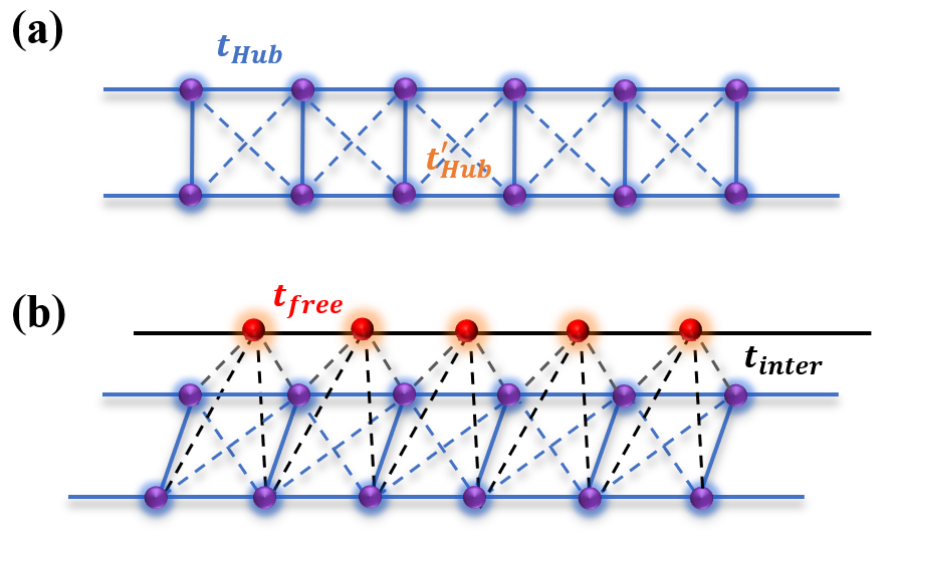}
\caption{Schematic illustration of the (a) two-legged Hubbard model and (b)
three-legged nickelate model.}
\label{ladder}
\end{figure}

The existence of competing or intertwining orders is common in cuprates\cite%
{RevModPhys.87.457}. The CO was observed in many families of cuprates with
or without accompanying spin order\cite{doi:10.1080/00018732.2021.1935698}.
Theoretically, the Hubbard model\cite%
{Hubbard238,doi:10.1146/annurev-conmatphys-090921-033948,doi:10.1146/annurev-conmatphys-031620-102024}
is believed to contain the microscopic ingredients of unconventional
superconductivity, endowed with a multitude of competing ground-state orders%
\cite{RevModPhys.84.1383}. Though recent result indicates extra terms are
needed to account for the superconductivity in the pure \cite{foot1} two-dimensional Hubbard
model \cite{PhysRevX.10.031016}, vertical stripe order is indeed established
in its ground state\cite{doi:10.1126/science.aam7127}. 
In a recent work, the instability of the CO in nickelates was studied in the
DFT+DMFT perspective \cite{2022arXiv220606985S}. In this work, to explore the
microscopic origin of the CO observed in the undoped parent
compound of nickelates, we study a minimal model (we name it nickelate model in this work) in which a
two-legged Hubbard ladder describes the Ni $3d_{x^2-y^2}$ electrons and a
free electron chain denotes the conduction electrons from the rare-earth
ions. As a comparison, we also study a pure \cite{foot2} two-legged Hubbard ladder with
hole doping. These two models are depicted in Fig.~\ref{ladder}. 

In this work, we employ the density matrix renormalization group (DMRG)
method\cite{PhysRevLett.69.2863,PhysRevB.48.10345}, which can provide highly
accurate results for the ground state of the nickelate model. When the
doping level in the Hubbard chains is tuned to $1/3$, a long-range CO with
period $3$ is found in the Hubbard chains of the three-legged nickelate model.
The electron pair-pair correlation decays algebraically with the distance between two
pairs. The spin density shows an exponentially decay but has an extremely
large correlation length, $i.e.$, larger than 20 lattice constants. We also
calculate the pure two-legged Hubbard ladder with $1/3$ hole doping, in
which a quasi-long range CO is present instead.
The comparison
indicates the self-doping effect can further enhance the CO, and the
Mott-Hubbard type model is relevant to the nickelate superconductors.
Similar numerical investigation was carried out for a two-band model on
four-legged cylinder\cite{HC-Jiang} {which focused on the much smaller
self-doping region.}
We also perform calculation on wider systems and find similar CO near $1/3$ self
doping but the spin
correlation changes to a perfect Neel type in the sense that there is no $\pi$-phase spin-flip in the
width 4 Hubbard and width 6 nickelate models. The details can be found in the Appendix.

\section{Models and Method}
The nickelate model we consider is a
three-legged model with the Hamiltonian
\begin{equation}
\begin{aligned} \hat{H} =&-\sum_{i, j \in H, \sigma} t_{i j}\left(\hat{d}_{i
\sigma}^{\dagger} \hat{d}_{j \sigma}+h.c.\right)+U \sum_{i \in H} \hat{n}_{i
\uparrow} \hat{n}_{i \downarrow} \\ &-\sum_{i \in H, k \in F, \sigma} t_{i
k}\left(\hat{d}_{i \sigma}^{\dagger} \hat{c}_{k \sigma}+ h.c.\right)
-\mu_{H} \sum_{i \in H} \hat{n}_{i} \\ &-\sum_{k, l \in F, \sigma} t_{k
l}\left(\hat{c}_{k \sigma}^{\dagger} \hat{c}_{l \sigma}+h.c.\right)-\mu_{F}
\sum_{k \in F} \hat{n}_{k} \end{aligned},  \label{ham}
\end{equation}
where $\hat{d}_{i \sigma}^{\dagger}(\hat{c}_{i \sigma}^{\dagger})$ is the Ni
$3d_{x^2-y^2}$ ($5d$ conduction) electron creation operator on site $%
i=(x_i,y_i)$ with spin $\sigma$, $\hat{n}_{i}=\sum_{\sigma} \hat{c}_{i
\sigma}^{\dagger} \hat{c}_{i \sigma}$ is the electron number operator, $%
\mu_H $ and $\mu_F$ are two chemical potentials, and $H$ and $F$ represent
the Hubbard and free chain, respectively. The nearest neighboring hopping
amplitude on the Hubbard chains is set as the energy unit, the next nearest
neighboring hopping is chosen as $t^{\prime}_{Hub}=0.1$, and the on-site
Hubbard interaction is set to $U=12$. The nearest neighboring hopping on the
free chain is set to $t_{free}=1.1$, and the hopping amplitude between the
Hubbard and free chains is chosen as $t_{inter}=0.2$. These chosen
parameters are relevant to the experimental materials.

\begin{figure}[tbp]
\includegraphics[width=0.45\textwidth]{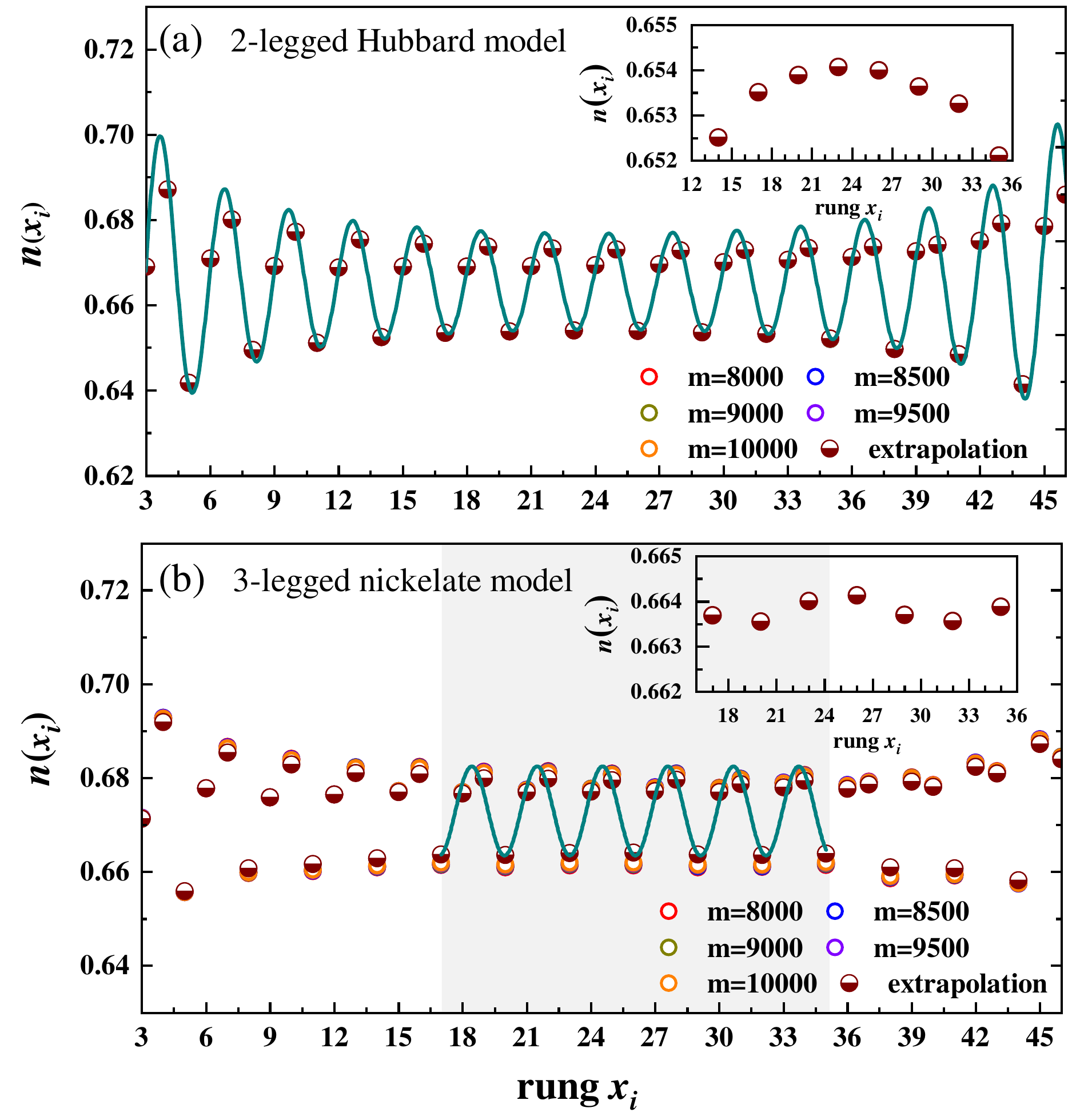}
\caption{Charge density profiles at $1/3$ hole doping for the (a) two-legged
Hubbard model and (b) three-legged nickelate model. The open circles
denote the local rung density for different number of kept states $m$ in
DMRG calculations and the half-filled one denote the extrapolated values.
The solid cyan line in (a) is
the fitting curve using Eq.~(\protect\ref{ext}). Fitting the results in (b)
with Eq.~(\protect\ref{ext}) gives both positive and negative $K_\protect\rho
$ depending on the range of data used, which indicates it actually
oscillates periodically without decay (the cyan line in (b) shows a
fit of the results using a cosine function in the middle of the system). The insets show the lower
peaks near the center of the systems.}
\label{2and3legElec}
\end{figure}

In our study, we assume that the whole system is kept at half-filling, $n ={%
N_e}/{N_{site}} = 1$, and we tune $\delta \mu=\mu_F-\mu_H$ to target the
desired hole (electron) doping level on the Hubbard (free) chain. The
average of hole concentration away from half-filling for each Hubbard chain
is defined as $\delta$, and the corresponding electron doping in the free
chain is $2\delta$. To measure the local spin density $\hat{s}_i=(\langle
\hat{n}_{i \uparrow}-\hat{n}_{i\downarrow}\rangle)/2$ instead of the more
demanding spin-spin correlation function, we apply a magnetic pinning field
with strength $h_m= 0.5 $ at the left edge of one Hubbard chain. 
{We also study the same system without pinning field and find the CO is
unchanged in the bulk of the system. The details of the results can be found in the Appendix.}
We also
calculate the pair-pair correlation function $D(i,j) = \langle \hat{%
\Delta}_i^{\dagger}\hat{\Delta}_j\rangle$, where
\begin{equation}
\hat{\Delta}_i^{\dagger}=\hat{c}_{(i,1),\uparrow}^{\dagger}\hat{c}%
_{(i,2),\downarrow}^{\dagger}-\hat{c}_{(i,1), \downarrow}^{\dagger}\hat{c}%
_{(i,2),\uparrow}^{\dagger},
\end{equation}
which creates a singlet pair across the rung of the Hubbard ladder.

The method we employ is the DMRG \cite{PhysRevLett.69.2863,PhysRevB.48.10345}%
, which can provide accurate results for such narrow systems with large
enough bond dimension. With open boundary conditions, we focus on ladder
with length $L_x=48$ and tune $\delta \mu = 0.9$ to target $1/3$ doping on
the Hubbard chains, which is relevant to the experimental observation\cite%
{rossi2021broken,krieger2021charge,tam2021charge}. The number of kept states
in the DMRG calculations is as large as $m=10,000$, {which gives truncation errors of
$6.90 \times 10^{-9}$ and $8.92 \times 10^{-6}$ for the two-legged Hubbard and three-legged nickelate models.}
 For the pure two-legged
Hubbard model, the results are converged with kept state, while for the
three-legged nickelate model, the free electron chain makes the entanglement
larger in the ground state and an extrapolation with truncation errors is
also performed to remove the finite kept state effect.

\begin{figure}[tbp]
\includegraphics[width=0.45\textwidth]{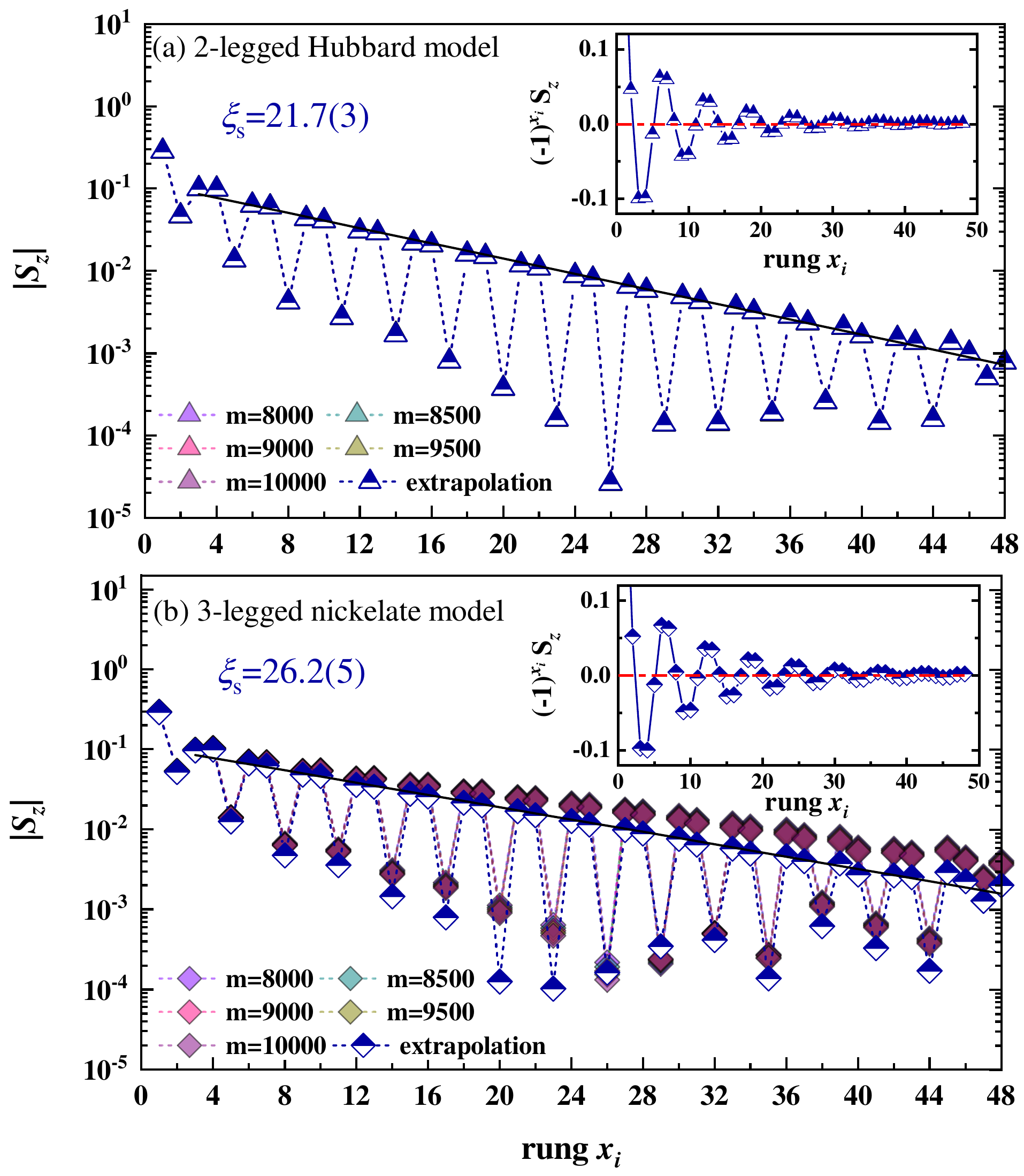}
\caption{Absolute values of the spin density on the (a) two-legged Hubbard
model and (b) three-legged nickelate model. Both results with finite kept
states $m$ (solid markers) and from an extrapolation with truncation errors
(half-solid markers) are shown. The insets display the staggered spin
density, where the dash-dotted horizontal lines represent zero.}
\label{2and3Sz}
\end{figure}

\section{Numerical Results}
\subsection{Hubbard ladders}
In Fig.~\ref{2and3legElec}(a), we show the
charge density profiles for the two-legged ladder with $\delta=1/3$. The
rung density is defined by $n(x_i)=\sum_{y=1}^{L_{y}}\left\langle {\hat{n}}%
_{i}(x,y)\right\rangle / L_{y}$ on the Hubbard ladders. The charge density
from DMRG with kept states $m = 8000$ to $m = 10000$ and the result from an
extrapolation with truncation error are on top of each other, which means
the DMRG results in Fig.~\ref{2and3legElec} (a) are converged (the same is
true for all the other quantities discussed in the remaining of this work).
The charge distribution displays a $\lambda_\rho=1/\delta$ periodic
structure. The vertical CO at long distances is dominated by a power-law
decay with the Luttinger exponent $K_\rho$, which can be fitted by the
Friedel oscillations induced by the boundaries\cite{PhysRevB.65.165122}
\begin{equation}
n(x) \approx A \frac{\cos \left(2 \pi N_{h} x / L+\phi_{1}\right)}{\left[L
\sin \left(\pi x / L+\phi_{2}\right)\right]^{K_{\rho} / 2}}+n_{0},
\label{ext}
\end{equation}
where $A$ is the amplitude, $\phi_1$ and $\phi_2$ are phase shifts, and $%
n_0=1-\delta$ is the mean charge density. Alternatively, we can obtain the
parameter $K_\rho$ from the finite-size scaling \cite{PhysRevB.92.195139},
because the density scales as
\begin{equation}
\delta n(L/2)=n(L/2)-n_{0}\sim L^{-K_{\rho}/2},
\end{equation}
around the center of the system. All fitting procedures give the similar
values of exponent $K_\rho$, which is close to 1.21(6). {The details of the
fittings can be found in the Appendix.}

\begin{figure}[tbp]
\includegraphics[width=0.45\textwidth]{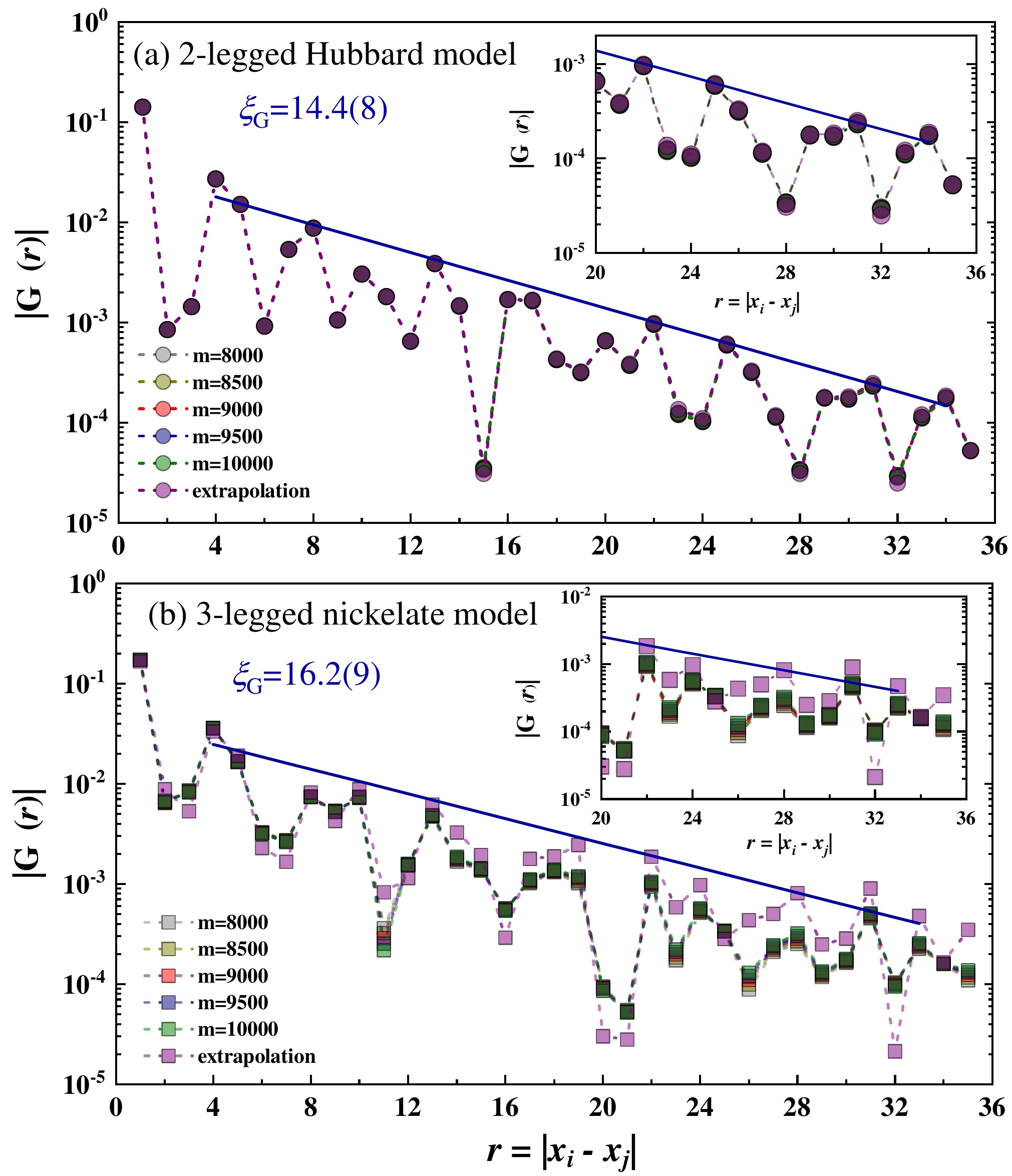}
\caption{Single-particle Green's function on the (a) two-legged Hubbard
model and (b) three-legged nickelate model. {In the calculation of $G_\sigma(r)$, we choose the 13th site on one Hubbard chain as the reference.} 
Both numerical results for finite kept states $m$ and extrapolated results are shown. In the insets,
the long-distance behavior is zoomed in. }
\label{2and3GF}
\end{figure}

The rung density $n(x)$ on the three-legged nickelate model from DMRG with $%
m = 8000$ to $m = 10000$ are shown in Fig.~\ref{2and3legElec} (b). Since the
three-legged system has a larger truncation error than the two-legged
Hubbard model in the DMRG calculation, we also perform an extrapolation with
truncation error $\epsilon $ to remove the finite kept state effect.

{A fit
following the same procedure as in the two-legged Hubbard model for the CO
correlation on the three-legged nickelate model gives both positive and negative
$K_\rho$ depending on the range of data used, which indicates it
actually oscillates periodically without decay, or $K_\rho \approx 0$,
suggesting the CO is actually a long-range one in the three-legged
nickelate model. {The details can be found in the Appendix.}}

The insets in Fig.~\ref{2and3legElec} show the lower peaks near the center
of the systems. In the inset of Fig.~\ref{2and3legElec} (a), we find a decay
from the center behavior which shows the envelope of the oscillation of
density governed by Eq. (\ref{ext}), while in the inset of Fig.~\ref%
{2and3legElec} (b), we find that the lower peak of density oscillates
without decay near the center which is the hallmark of the existence of true
long-range CO in the nickelate model. Comparing the CO in the three legged
nickelate model and the pure two-legged Hubbard model, we conclude that the
free electron chain, which plays the role of a charge reservoir, can enhance
the CO in the nickelate model. {We also observe an induced CO in the free
chain with period $6$ in the nickelate model as will be shown later.}

In Fig.~\ref{2and3Sz}, we display the absolute value of magnetization $%
\langle \hat{s}_{i}^z \rangle$ with different number of kept states $m$ in
DMRG calculations as well as the extrapolated result. For both systems, $%
\langle \hat{s}_{i}^z \rangle$ displays an exponential decay as $\langle
\hat{s}_{i}^z \rangle \propto \mathrm{e}^{-x_i / \xi_{\mathrm{s}}}$ at long
distances, with a finite correlation length $\xi_{\mathrm{s}}$. Comparing
with the two-legged Hubbard model, the three-legged nickelate model has a
slightly larger $\xi_{\mathrm{s}}$, implying a smaller spin excitation gap.
The staggered spin density $(-1)^{x_i} \langle \hat{s}_{i}^z \rangle$ in the
insets of Fig.~\ref{2and3Sz} show a spatial modulation with a wavelength
twice that of the charge density, and have a $\pi$ phase flip at the
hole-concentrated sites, which is the feature of the so-called stripe order%
\cite{PhysRevB.40.7391,doi:10.1143/JPSJ.59.1047,PhysRevB.39.9749,refId0}.

The single-particle Green's function of the electrons on the 
Hubbard ladder is defined by 
\begin{equation}
G_\sigma(r)=\langle \hat{c}_{(x_0,y),\sigma}^\dagger\hat{c}_{(x_0+r,y),\sigma}\rangle,
\end{equation}
which involves both charge and spin degrees of freedom. The corresponding 
numerical result has shown in Fig.~\ref{2and3GF}. Due to the presence of 
the small energy gap in the spin excitations, this electron Green's function 
is found decaying exponentially as $G_\sigma(r)\propto\mathrm{e}^{-r/\xi_{\mathrm{G}}}$, 
from which a correlation length can be extracted as $\xi_{\mathrm{G}}=14.4(8)$ 
and $16.2(9)$ for the two-legged Hubbard model and three-legged nickelate model, 
respectively. 

\begin{figure}[tbp]
\includegraphics[width=0.45\textwidth]{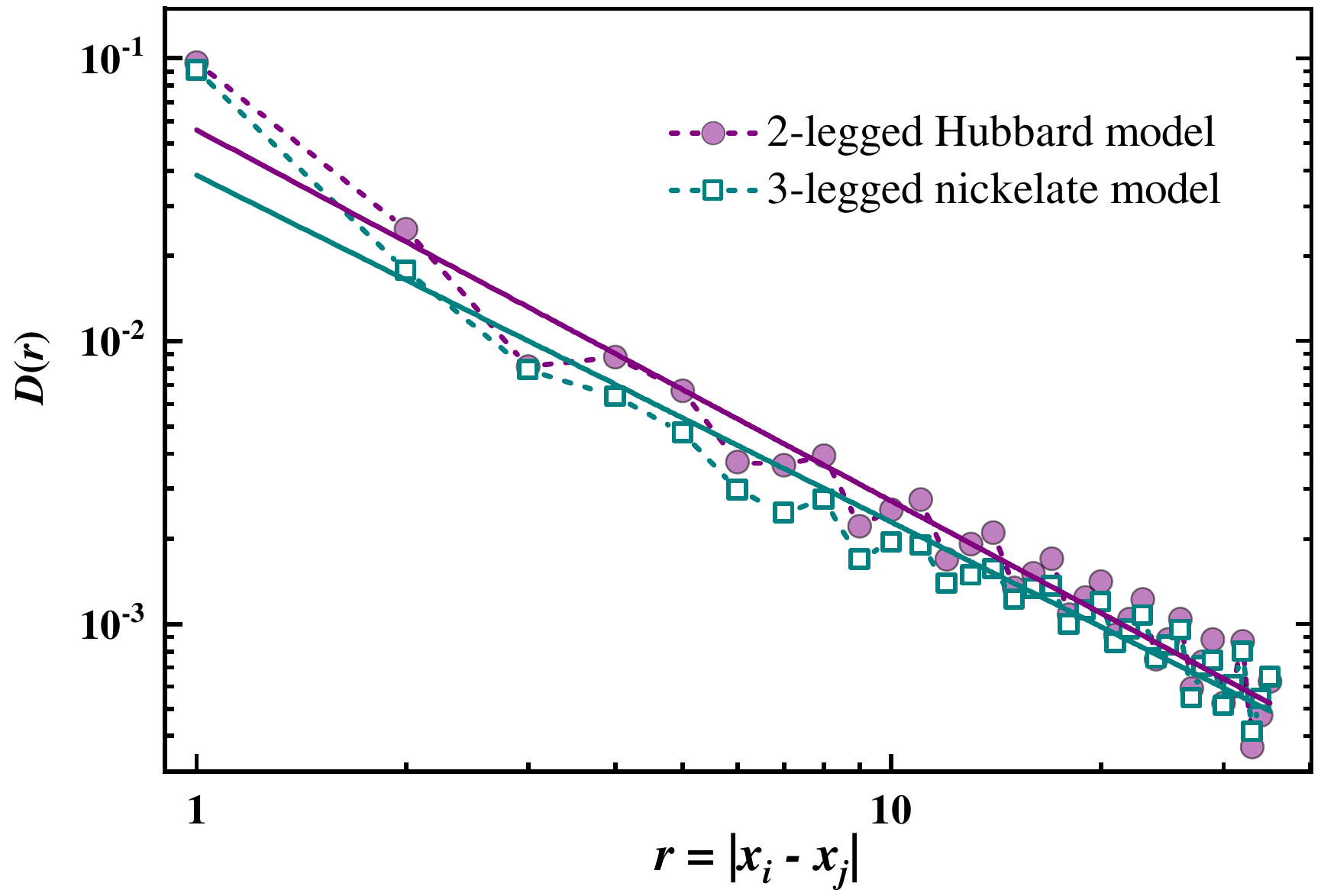}
\caption{Pair-pair correlation function for the two-legged Hubbard model
(solid circle) and three-legged nickelate model (empty square) at $1/3$
doping. {In the calculation of $D(r)$, we choose the 13th vertical bond as the reference.} The power-law fits $D(r)\propto r^{-K_{sc}}$ for both systems give
close exponents $K_{sc}$. Only extrapolated results with truncation errors
are shown. {Results with finite kept state and the extrapolation can be found
in the Appendix.} }
\label{2and3PairCorr}
\end{figure}

More importantly, the singlet pair-pair correlation function $D(i,j)$ for the
pure two-legged Hubbard model and the three-legged nickelate model are also 
calculated and displayed in Fig.~\ref{2and3PairCorr}. The given numerical results 
are from the extrapolation with truncation error in DMRG calculations. {The details
of extrapolation are provided in the Appendix.} The singlet pair-pair correlation
function decays algebraically for both systems, $i.e.$, $D(r)\propto
r^{-K_{sc}}$ with $K_{sc}$=$1.3(1)$ and $K_{sc} =1.2(1)$ for the pure
two-legged Hubbard model and the three-legged nickelate model, respectively.
The algebraic decay of pairing correlation indicates the instability of
superconductivity in both systems. 

Finally we summarize the exponents and correlation lengths for the charge,
spin, single-particle, and singlet pairing of the correlation functions for 
both models with $1/3$ hole doping in Table.~\ref{fit_parameters}. Comparing 
the three-legged nickelate and the two-legged Hubbard model, the major difference 
is the enhancement of charge order by including the charge reservoir, while 
the pairing, spin, and single particle properties are very similar. Therefore,
within the simplified model Hamiltonian, the present numerical results have 
clearly demonstrated their similarities between the nickelates and cuprates. 

\begin{table}[tbp]
\caption{List of the extracted parameters for charge, spin, pair-pair
correlations, and single-particle Green's function. Both the two-legged
Hubbard model and the three-legged nickelate model results are shown for
comparison.}
\includegraphics[width=0.45\textwidth]{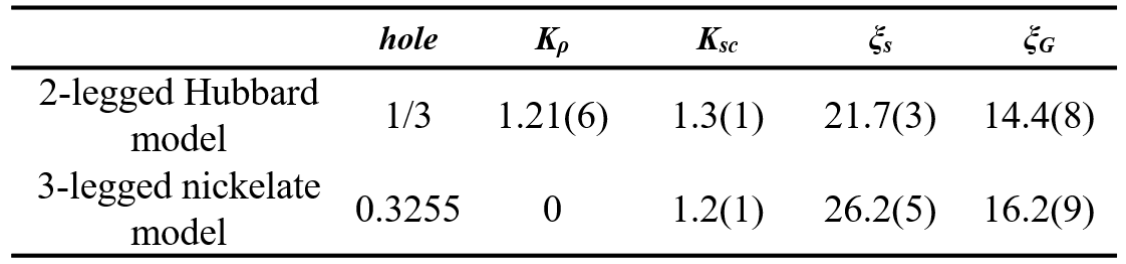}
\label{fit_parameters}
\end{table}

\subsection{Free chain in the three-legged nickelate model}

\begin{figure*}[t]
   \includegraphics[width=0.8\textwidth]{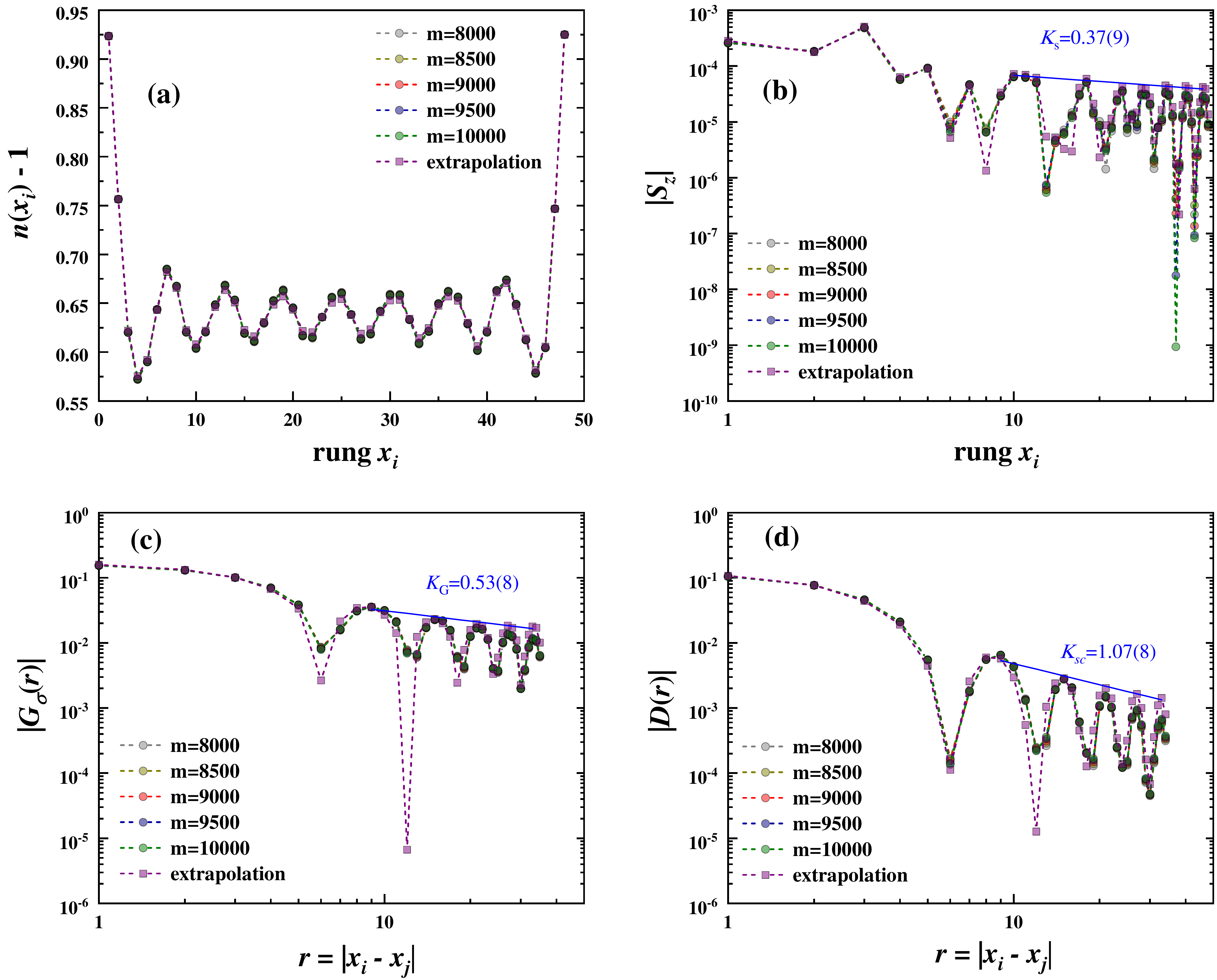}
   \caption{(a) local charge density, (b) local spin density, (c) single-particle Green's function, and (d) pair-pair correlation on the free chain of the three-legged nickelate model. 
    {The reference bond (site) is set at the 13th horizontal bond (site) when calculating $D(r)$ ($G_\sigma(r)$)}. Solid lines are
   	algebraic fits and the corresponding exponents are listed. The induced charge order in the free chain has a period of $6$ lattice constants.}
   \label{freechain_appendix5}
\end{figure*}

In Fig.~\ref{freechain_appendix5}, we show the local charge density, local spin density, single-particle
Green's function, and pair-pair correlation for the free chain in the three-legged nickelate model. Because the whole system is set at half-filling and the Hubbard chains of the three-legged nickelate model are about $1/3$ hole-doped, the free chain is electron doped (with averaged density about $5/3$). We
can find an induced charge order with period $6$ in the free chain, while the induced spin density is
negligibly small due to the absence of the Hubbard interaction.  We can also find an induced quasi-long
range like pair-pair correlation with the same period as the local charge density. 

\section{Summary and Perspectives}
We have studied a three-legged nickelate
model which {might be} relevant to the infinite-layer nickelate superconductors. Our
accurate DMRG calculations have yielded a long-range charge order with period $3$
in the undoped compound due to the presence of self-doping effect from the
rare-earth-layer $5d$ itinerant electrons. Our results are consistent with
spectroscopic experiments\cite{krieger2021charge,tam2021charge}, which
showed a charge order resulted from the strong correlation of Ni $3d$
orbitals and self-doping effect from the Nd $5d$ orbitals. Remarkably,
different from the quasi-long range charge order in the pure two-legged
Hubbard with $1/3$ hole doping, the charge order in three-legged nickelate
model is long-ranged instead, indicating that the charge reservoir of the
free chain can enhance the charge order in the Hubbard ladder. At the same
time, a small energy gap is present in the spin excitations, leading to an
extremely large correlation length. 

A quasi-long-range pairing
correlation is also determined in the nickelate model, which might be a
precursor of superconducting order in two dimension case with tuned
parameters. Although we focus on the parent compound in this work, it is
usually believed that the long-range charge order disfavors the
superconductivity. Experimentally, the charge order becomes weaker with the
increase of hole doping and finally disappears at $20\%$ hole doping, where
the superconductivity sets in\cite{krieger2021charge}. How the charge order
melts with hole doping in the nickelate model and the evolution of pairing
correlation with doping will be an interesting topic for future study.

\section{acknowledgments}
Y. Shen and M. P. Qin thank Weidong Luo for his generosity to provide
computational resources for this work. G. M. Zhang acknowledges the support
from the National Key Research and Development Program of MOST of China
(2017YFA0302902). M. P. Qin acknowledges the support
from the National Key Research and Development Program of MOST of China
(2022YFA1405400), the National Natural Science Foundation of China (Grant
No. 12274290) and the sponsorship from Yangyang Development Fund. All the DMRG calculations are carried out with iTensor library \cite{itensor}.

\bibliography{ladder_DMRG}

\begin{thebibliography}{40}
\bibitem[{foo({\natexlab{a}})}]{foot1}
\bibinfo{note}{Pure means only nearest neighbouring hoppings are considered in
  the Hubbard model.}
\bibitem[{foo({\natexlab{b}})}]{foot2}
\bibinfo{note}{Pure here means the free electron chain in the three-legged
  nickelate model is removed.}
\end{thebibliography}

\appendix
\section{Dependence of $K_\rho$ on the range of data}
In Table. \ref{K_cdw_range_appendix1}, we list the extracted $K_\rho$ and $n_0$
using Eq. (3) in the main text for the two-legged $1/3$ hole doped Hubbard model with
length $L = 48$. The values of $K_\rho$ are obtained from the fits by using charge density
from different range of sites. The results are consistent with the values in the main
text $K_\rho = 1.21 (6)$.

\begin{table}[b]
	\caption{The dependence of the extracted $K_\rho$ and $n_0$ (using Eq. (3) in the main text) on
		   the range of sites used in the fit for the two-legged Hubbard model ($L=48$).}
	\includegraphics[width=0.4\textwidth]{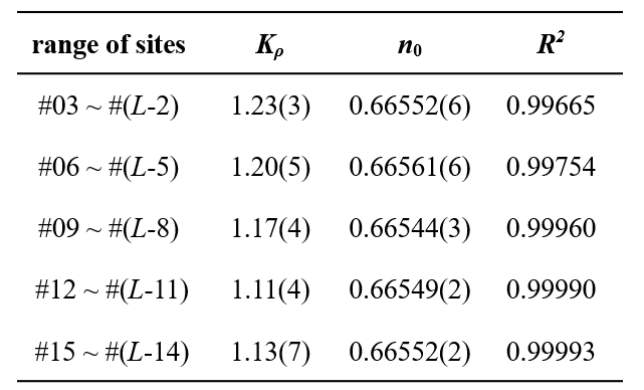}
	\label{K_cdw_range_appendix1}
\end{table}

\begin{figure}[b]
   \includegraphics[width=0.4\textwidth]{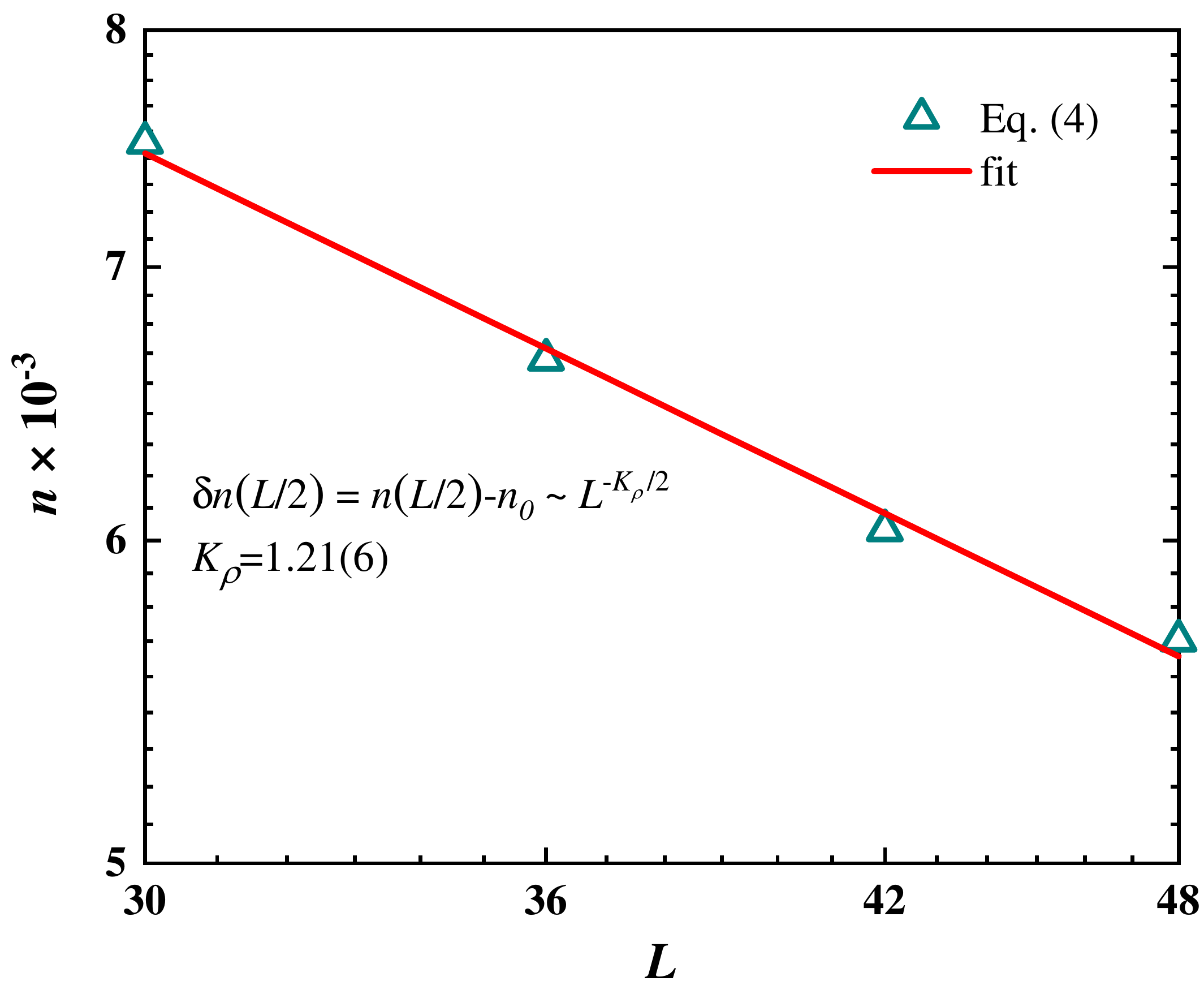}
   \caption{Finite-size scaling of $\delta n(L/2)$ as a function of the system size $L$ for the two-legged Hubbard model.}
   \label{K_cdw_scaling_appendix2}
\end{figure}

In Fig.~\ref{K_cdw_scaling_appendix2}, we show the fit using the value at the center
of the systems for different sizes following Eq. (4) in the main text. The extracted
$K_\rho$ is consistent with the values listed in Table. \ref{K_cdw_range_appendix1}.

{In Table.~\ref{K_cdw_range_appendix2}, we list the extracted $K_\rho$ and $n_0$
using Eq. (3) in the main text for the three-legged $1/3$ hole doped nickelate model with
length $L = 48$. The values of $K_\rho$ are obtained from the fits by using charge density
from different range of sites. We can find that the fit gives that both positive and negative
$K_\rho$ depending on the range of data used, which indicates it actually oscillates periodically
without decay, or $K_\rho \approx 0$, suggesting the CO is actually a long-range one
in the three-legged nickelate model.}

\begin{table}[t]
	\caption{The dependence of the extracted $K_\rho$ and $n_0$ (using Eq. (3) in the main text) on
		the range of sites used in the fit for the three-legged nickelate model ($L=48$).}
	\includegraphics[width=0.4\textwidth]{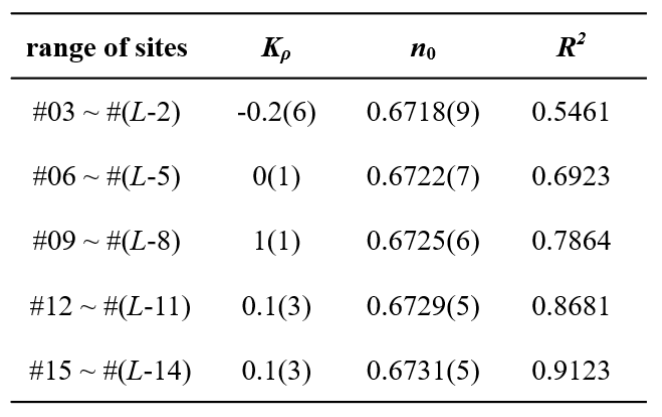}
	\label{K_cdw_range_appendix2}
\end{table}

\section{Extrapolation with truncation errors in DMRG}

\begin{figure*}[t]
   \includegraphics[width=0.8\textwidth]{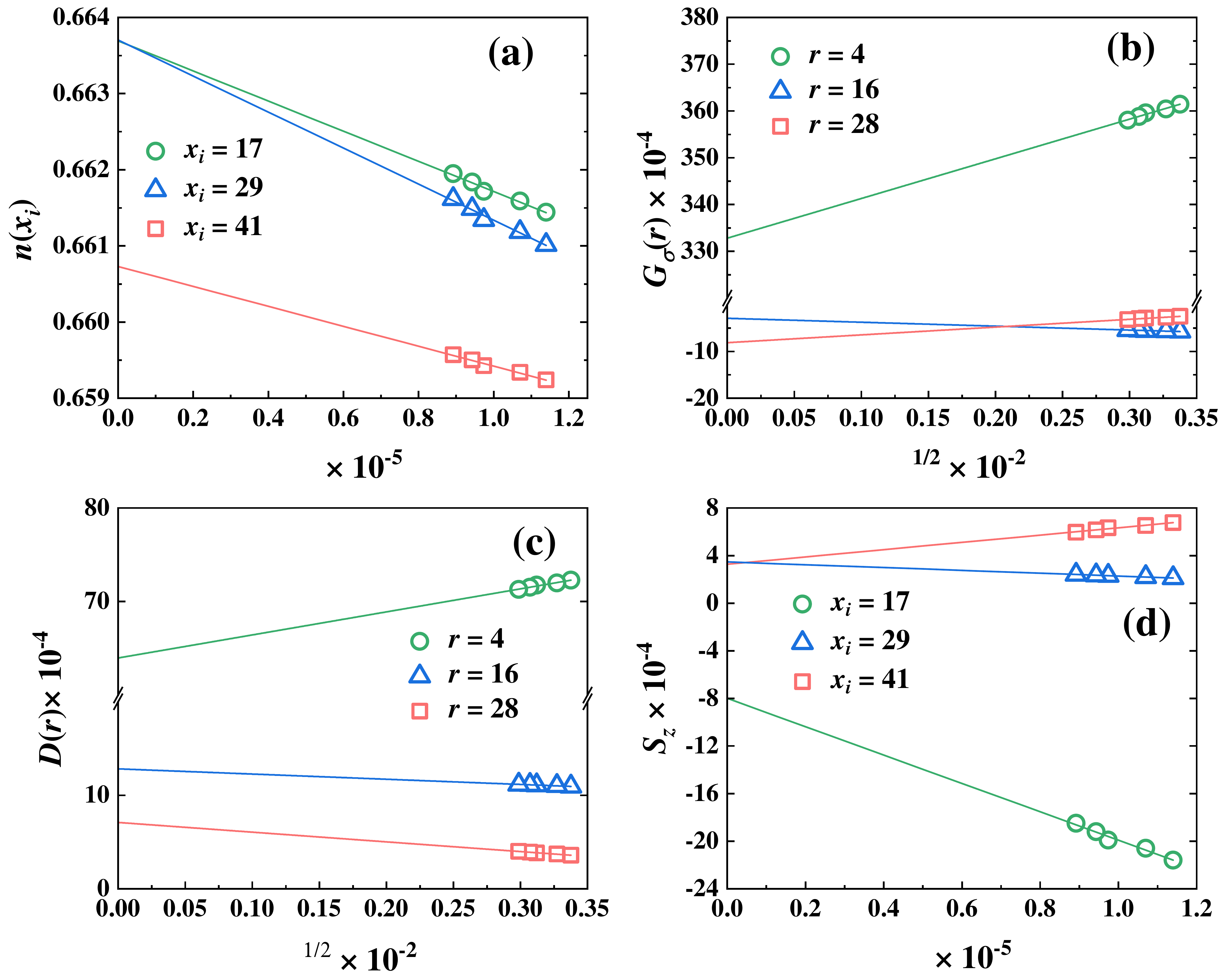}
   \caption{Extrapolation of observables of the three-legged nickelate model as a function of the truncated error $\epsilon$ in DMRG calculations. Representative examples of (a) the local rung density $n(x_i)$, (b) single-particle Green's function $G_\sigma(r)$, (c) pair-pair correlation function $D(r)$, and (d) local spin $\langle \hat{s}_{i}^z \rangle$ are shown. {The reference bond (site) is set at the 13th vertical bond (site on one Hubbard chain) when calculating $D(r)$ ($G_\sigma(r)$)}. Local observables such as charge density and spin density are fitted linearly with $\epsilon$, while correlation functions are fitted linearly with $\sqrt{\epsilon}$.}
   \label{fit_appendix4}
\end{figure*}

In Fig.~\ref{fit_appendix4}, we show the extrapolation of DMRG results for local
charge density, the single-particle Green's function, pair-pair correlation,
and local spin density with truncation errors. We choose three typical sites in
Fig.~\ref{fit_appendix4}. The local quantities (charge and spin) are extrapolated with
truncation error, while the correlations
(single-particle Green's function and pair-pair correlation) are extrapolated with the square root of the truncation error.

\section{Results without magnetic pinning field}
{We also perform calculation of the three-legged nickelate model without any magnetic pinning field. In the left panel of Fig.~\ref{no-pinning-conv} we show
how the local spin density vanishes with the the increase of bond dimension $m$. The local spin density is under $10^{-6}$ with bond dimension $m = 9000$ which
means the SU(2) symmetry is almost restored and the DMRG results are well converged. In the right panel of Fig.~\ref{no-pinning-conv} we show a comparison
of the charge density with and without pinning field on the left edge. We can find that the charge density remains almost unchanged in the
bulk, though the results in the left edge show discrepancy as expected.  
\begin{figure*}[t]
	\includegraphics[width=0.41\textwidth]{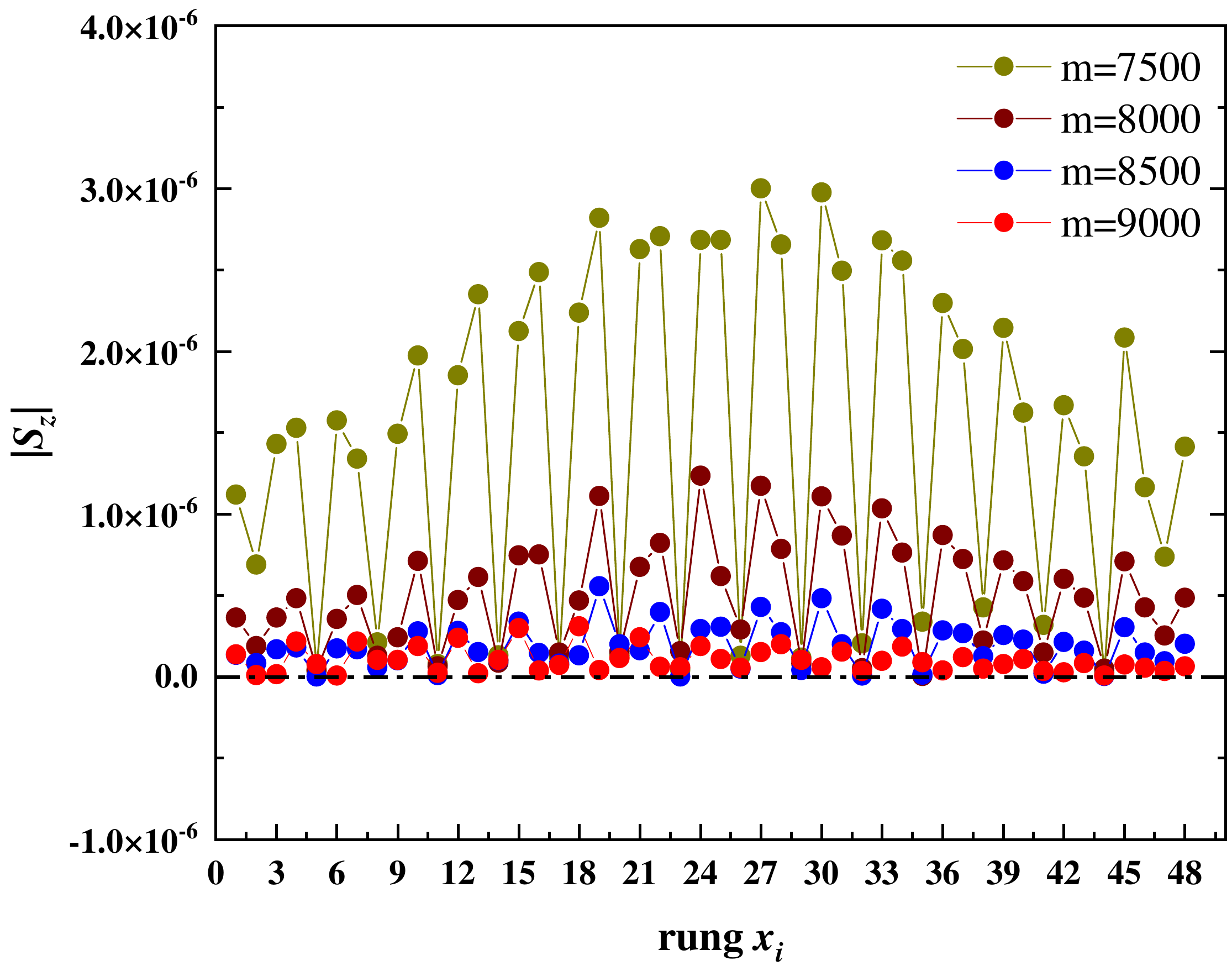}
	\includegraphics[width=0.4\textwidth]{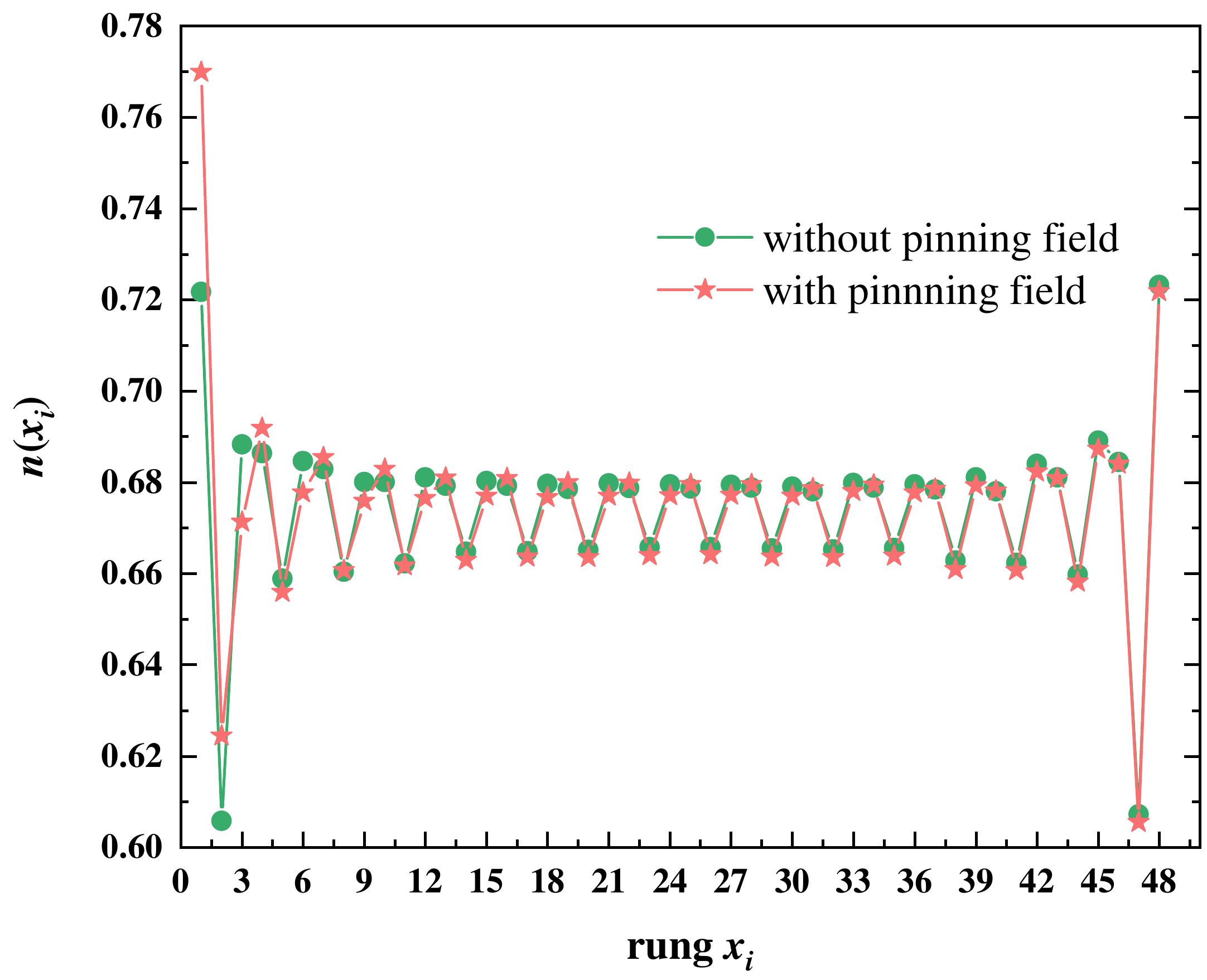}
	\caption{Left: the vanish of local spin density with the increase of bond dimension $m$ for the three-legged
		nickelate model without magnetic pinning field. Right: comparison of the charge density for system
	with and without edge pinning field for the two-legged Hubbard model.}
	\label{no-pinning-conv}
\end{figure*}
}

\section{Results for width-4 and width-6 systems}
{
We also study the Hubbard and nickelate models on width 4 and 6 ladders. The charge, spin and pairing correlation are shown in Fig.~\ref{width-4}.
The period 3 CO is also found in these wider system. But different from the 2-legged Hubbard and 3-legged nickelate models, there is no
$\pi$ phase flip in the spin order which means the spin correlation is perfect Neel type. But we can still find a similar period
$6$ modulation in the spin density. The result in the right panel of Fig.~\ref{width-4} shows pairing correlation is suppressed in wider system.     
\begin{figure*}[t]
	\includegraphics[width=0.3\textwidth]{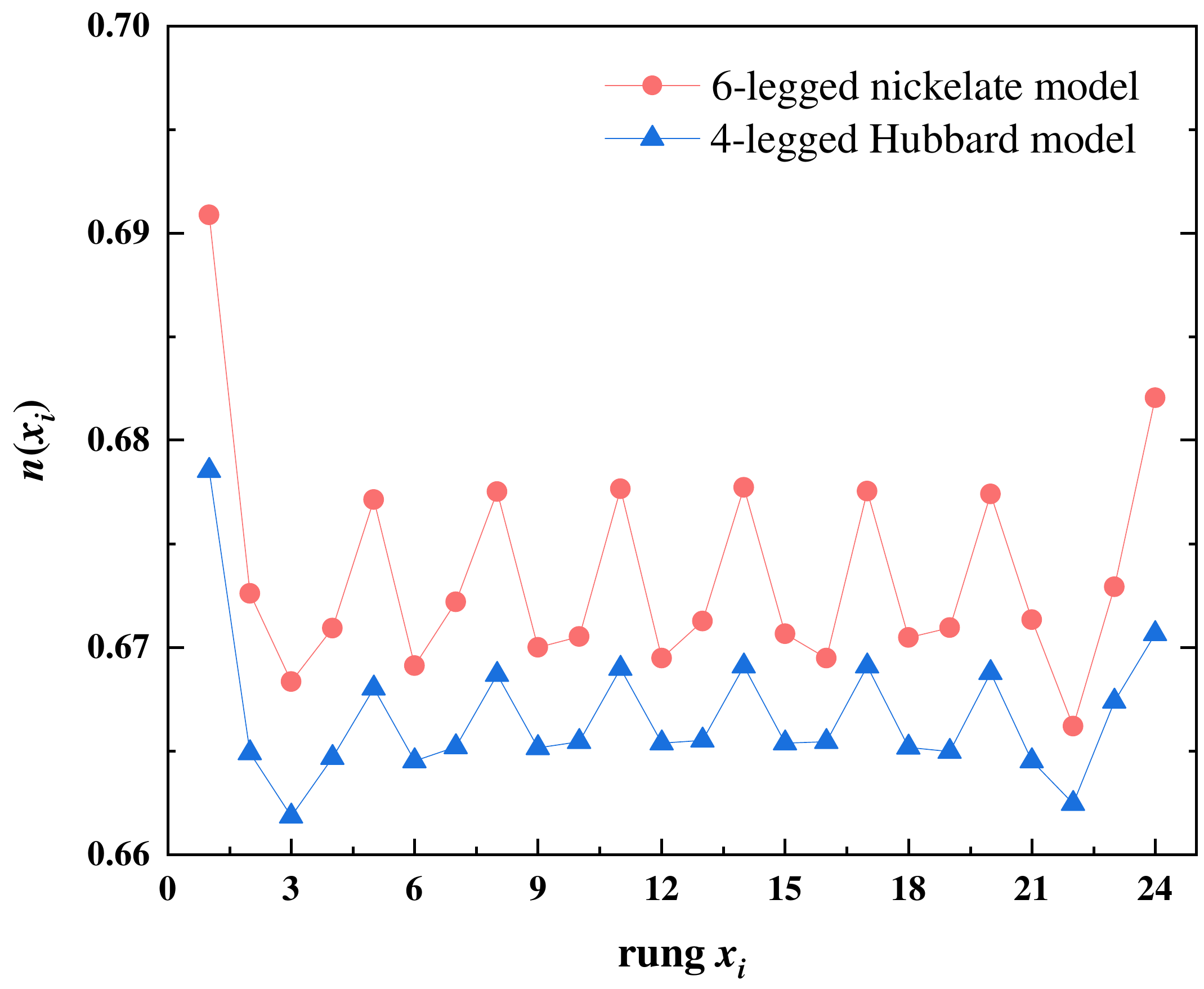}
	\includegraphics[width=0.3\textwidth]{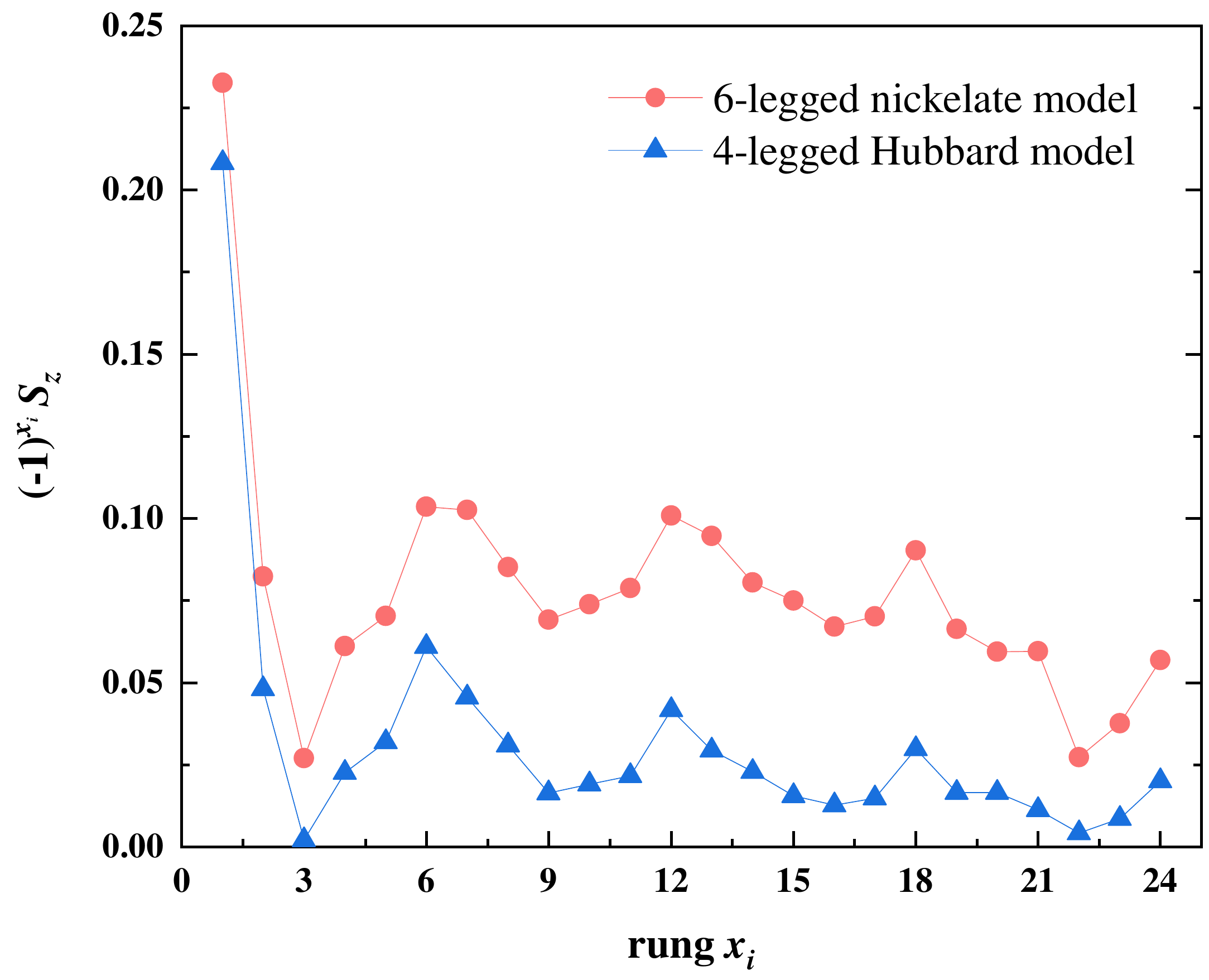}
	\includegraphics[width=0.3\textwidth]{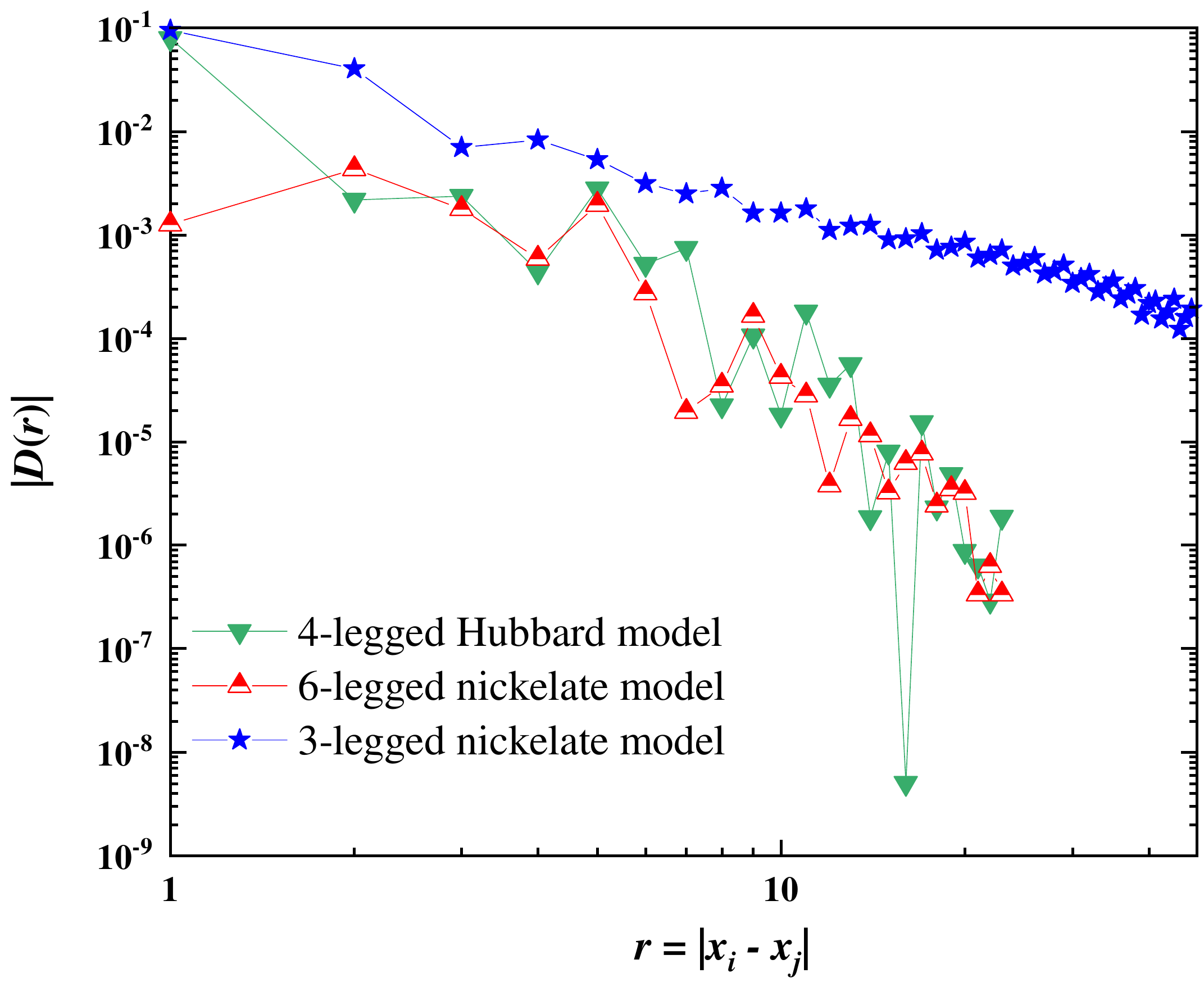}
	\caption{Charge density (left), spin density (middle), and pair-pair correlation (right) for the 4-legged Hubbard and 6-legged nickelate models. In the left panel, we can find that
	the period 3 CO remains the same as the 2-legged and 3-legged systems. In the middle panel, the spin changes to Neel antiferromagnetic type without $\pi$ phase flip found in the 2-legged and 3-legged systems. In the right panel, the pairing
    is found to be suppressed comparing to 3-legged systems.}
	\label{width-4}
\end{figure*}
}




\end{document}